\documentclass[12pt]{iopart}
\usepackage{iopams,setstack}
\begin{document}

\title[New Symmetries and Multi-Hamiltonian Structures for the Toda Lattice]{New non-Noetherian Symmetries and Multi-Hamiltonian Structures for the Toda Lattice}

\author{Felipe A Asenjo$^1$, Sergio A Hojman$^2$}

\address{$^1$  Departamento de F\'{\i}sica, Facultad de Ciencias, Universidad de Chile, Las Palmeras 1356, \~ Nu\~ noa, Santiago, Chile}\ead{fasenjo@zeth.ciencias.uchile.cl}
\address{$^2$ Departamento de Ciencias, Facultad de Artes Liberales, Universidad Adolfo Ib\'a\~nez, Diagonal Las Torres 2640, Pe\~nalol\'en, Santiago, Chile}\ead{sergio.hojman@uai.cl}

\begin{abstract}

New symmetry transformations for the n-dimensional Toda lattice are
presented. Their existence allows for the construction of several
first order Lagrangian structures associated to them. The
multi-Hamiltonian structures are derived from Lagrangians in detail.
The set of symmetries generates a Lie algebra.
\end{abstract}



\section{Introduction}

The Toda chain \cite{toda0, toda} is a well known non trivial
exactly solvable model which considers a lattice where each site
interacts with its nearest neighbors. This lattice has been very
extensively studied \cite{flaschka}, using different approaches in
Lie algebras \cite{bogo},  as well as in quantum \cite{rod, hader,
ikeda}, and in relativistic systems \cite{ruij, suris, nunes}.
Besides these studies, other generalizations have been carried out
\cite{mikha, fahmi}.

In this paper, we focus in the first order approach for the Toda
lattice and show its rich structures. Recently, Chavchadnize
\cite{chav} has found a new symmetry transformation for the two
dimensional Toda lattice. In this note, we present an extension of
this symmetry transformation to $n$-dimensional lattices as well as
four new symmetry transformations. Using these symmetries we
construct new different Lagrangian structures for the $n$
dimensional Toda model. Each one of these structures gives rise to
new Hamiltonian functions and different (but equivalent) first order
Euler Lagrange equations. Moreover, the new set of symmetry
transformations presented here generates a Lie algebra.

\section{Definitions for the Toda Lattice. First and second order formalisms.}

The n dimensional  Toda lattice may be described by the second order Lagrangian
$L_{(2)}$ which is a function of $n$ independent variables $q^i$ and their
time derivatives ($i,j = 1,........,n$)
\begin{equation}
L_{(2)}(q^i, \dot{q}^j)=\frac{1}{2}\sum_{k=1}^{n}(\dot{q}^k)^2-\sum_{k=1}^{n-1} e^{q^k-q^{k+1}}\, ,
\label{solag}
\end{equation}
which gives rise to the following $n$ second order Euler--Lagrange
equations
\begin{equation}
\ddot{q}^k-e^{(q^{k-1}-q^k)}+e^{(q^k-q^{k+1})}=0\, ,
\label{ec1}
\end{equation}
where $k = 1,......,n$ and the conventions
\begin{equation}\label{conv}
\eqalign{e^{q^0-q^1}&\equiv0\, ,\\
e^{q^n-q^{(n+1)}}&\equiv0\, ,
}\end{equation}
have been used.

The Hamiltonian structure associated to the Lagrangian formulation
(\ref{solag}) is defined by
\begin{equation}
H(q^i, p_j)=\frac{1}{2}\sum_{k=1}^{n}(p_k)^2+\sum_{k=1}^{n-1}
e^{q^k-q^{k+1}}\, ,
\label{ec2}
\end{equation}
with $i,j = 1,........,n\ $ and where the momenta $p_j$ are defined as usual by
\begin{equation}
p_j\equiv \frac{\partial L}{\partial \dot{q}^j}= \dot{q}^j\, .
\label{ec10}
\end{equation}

Let us define $2n$ variables by
\numparts
\begin{eqnarray}\label{eqaaa1}
x^i&=q^i\, ,\label{eqaaa1A}\\
x^{n+j}&=p_j\, .\label{eqaaa1B}
\end{eqnarray}
\endnumparts

Thus, the Toda model Hamiltonian \eref{ec2} can be written as
\begin{equation}
H=\frac{1}{2}\sum_{j=1}^n(x^{n+j})^2+\sum_{i=1}^{n-1}e^{x^i-x^{i+1}}\, ,
\label{hamil1}
\end{equation}
and the Toda model first order Lagrangian is
\begin{equation}
L_{(1)}=\sum_{i=1}^nx^{n+i}{\dot
x}^i-\frac{1}{2}\sum_{i=1}^n(x^{n+i})^2-\sum_{i=1}^{n-1}e^{x^i-x^{i+1}}\, .
\label{lagra1}
\end{equation}

This Lagrangian gives rise to $2n$ first order equations
\numparts
\begin{eqnarray}\label{eqmotion2n}
{\dot x}^i&=x^{n+i}\label{eqmotion2nA}\\
{\dot x}^{n+j}&=e^{(x^{j-1}-x^j)}-e^{(x^j-x^{j+1})}\label{eqmotion2nB}
\end{eqnarray}
\endnumparts
where $i,j =1,...,n$, and the conventions
\begin{equation}\label{conventions}
\eqalign{e^{x^0-x^1}&\equiv0\\
e^{x^n-x^{(n+1)}}&\equiv0
}\end{equation}
are used.

These $2n$ equations are equivalent to the previous $n$ second order \Eref{ec1}. These first order equations can be written as ($a,b=1,.....,2n$)
\begin{equation}
{\dot x}^a=f^a(x^{b}, t)\, ,
\label{eqmot1}
\end{equation}
and
\begin{equation}\label{f}
\eqalign{f^j&=x^{n+j}\, ,\\
f^{n+j}&=e^{(x^{j-1}-x^j)}-e^{(x^j-x^{j+1})}\, ,
}\end{equation}
with $j=1,...,n$ and where the previous conventions apply.

A symmetry transformation for a system of differential equations is
defined by an infinitesimal transformation of the variables $x^{a}$
\begin{equation}
x'^{a}= x^{a}+\epsilon \eta^{a}(x^{b}, t)\, , \label{symm1}
\end{equation}
such that $x'^{a}$ satisfies \Eref{eqmot1} if $x^{a}$ does,
i.e.,
\begin{equation}
{\dot x}'^a=f^a(x'^{b}, t)\, .
\end{equation}

Therefore, the vector $\eta^{a}(x^{b}, t)$  satisfies
\begin{equation}
\frac{\partial \eta^{a}(x^{b}, t)}{\partial t}+ \frac{\partial
\eta^{a}(x^{b}, t)}{\partial x^{c}}f^{c}- \frac{\partial
f^{a}(x^{b}, t)}{\partial x^{c}}\eta^{c}=0 \label{symm2}
\end{equation}
up to first order in $\epsilon$. Note that these symmetries fulfill ($j=1,...,n$)
\begin{equation}
\eta^{n+j}=\frac{\bar d\eta^{j}}{dt}=\frac{\partial\eta^j}{\partial x^a}f^a+\frac{\partial \eta^j}{\partial t}\, .
\end{equation}

The \Eref{symm2} is equivalent to the so called Master Equation \cite{sah1}
\begin{equation}
\left(\frac{\partial}{\partial t}+\underset{f}{{\cal
L}}\right)\eta^a=0\, ,
\end{equation}
where $\underset{f}{\cal L}$ is the Lie derivative along the vector
$f^{a}$ \cite{sah2}, which, for a vector $\eta^a$ may be written as
\begin{equation}
\underset{f}{\cal L}\ \eta^a=\frac{\partial\eta^a}{\partial x^b}f^b-\eta^b\frac{\partial f^a}{\partial x^b}\, .
\end{equation}

Note that $\underset{f}{\cal L}\ \eta^a=-\underset{\eta}{\cal L}\ f^a$.

%%%%%%%%%%%%%%%%%%%%%%%%%%%%%%%%%%%%%%%%%%%%%%

\section{New Symmetry Transformations for the Toda Lattice}

We will exhibit five different solutions to \Eref{symm2} for the
Toda Model, with $f^{a}$ given in \eref{f}. These symmetry
vectors give rise to a Lie algebra where the commutation operation
is defined by the Lie derivative. Further details are given below.

Chavchanidze \cite{chav} found a non trivial symmetry vector for the
two dimensional Toda lattice. In fact, Chavchanidze showed that the
vector $\eta_{(1)}=(\eta^1_{(1)}, \eta^2_{(1)}, \eta^3_{(1)},
\eta^4_{(1)})$ with
\begin{equation}
\eqalign{\eta^1_{(1)}&=2x^3+\frac{1}{2}x^4+\frac{t}{2}\left(\left(x^3\right)^2+e^{(x^1-x^2)}\right)\, ,\\
\eta^2_{(1)}&=x^4-\frac{1}{2}x^3+\frac{t}{2}\left(\left(x^4\right)^2+e^{(x^1-x^2)}\right)\, ,\\
\eta^3_{(1)}&=\frac{1}{2}\left(x^3\right)^2-e^{(x^1-x^2)}-\frac{t}{2}\left(x^3+x^4\right)e^{(x^1-x^2)}\, ,\\
\eta^4_{(1)}&=\frac{1}{2}\left(x^4\right)^2+2e^{(x^1-x^2)}+\frac{t}{2}\left(x^3+x^4\right)e^{(x^1-x^2)}\,
, } \label{chavsym2d}
\end{equation}
is a symmetry vector for \Eref{eqmot1} with $n=2$, i.e., the
symmetry vector satisfies \Eref{symm2}.

For $n=3$ (a slightly incorrect version of this symmetry
transformation appears in  \cite{chav}), it can be shown that
$\eta_{(1)}=(\eta^1_{(1)}, \eta^2_{(1)}, \eta^3_{(1)},
\eta^4_{(1)},\eta^5_{(1)}, \eta^6_{(1)})$ is a symmetry vector for
\Eref{eqmot1} with
\begin{equation}
\eqalign{\eta^1_{(1)}&=3x^4+\frac{1}{2}x^5+\frac{1}{2}x^6+\frac{t}{2}\left(\left(x^4\right)^2+e^{(x^1-x^2)}\right)\, ,\\
\eta^2_{(1)}&=2x^5-\frac{1}{2}x^4+\frac{1}{2}x^6+\frac{t}{2}\left(\left(x^5\right)^2+e^{(x^1-x^2)}+e^{(x^2-x^3)}\right)\, ,\\
\eta^3_{(1)}&=x^6-\frac{1}{2}x^4-\frac{1}{2}x^5+\frac{t}{2}\left(\left(x^6\right)^2+e^{(x^2-x^3)}\right)\, ,\\
\eta^4_{(1)}&=\frac{1}{2}\left(x^4\right)^2-2e^{(x^1-x^2)}-\frac{t}{2}\left(x^4+x^5\right)e^{(x^1-x^2)}\, ,\\
\eta^5_{(1)}&=\frac{1}{2}\left(x^5\right)^2+3e^{(x^1-x^2)}-e^{(x^2-x^3)}+\frac{t}{2}\left(x^4+x^5\right)e^{(x^1-x^2)}\nonumber\\
&\quad-\frac{t}{2}\left(x^5+x^6\right)e^{(x^2-x^3)}\, ,\\
\eta^6_{(1)}&=\frac{1}{2}\left(x^6\right)^2+2e^{(x^2-x^3)}+\frac{t}{2}\left(x^5+x^6\right)e^{(x^2-x^3)}\, .
}\end{equation}

We have generalized Chavchanidze's result for the n dimensional Toda
lattice symmetry vector $\eta_{(1)}=(\eta^j_{(1)},\eta^{n+j}_{(1)})$
with $j=1,.....,n$ as follows
\begin{equation}\label{eta1}
\eqalign{\eta^j_{(1)}&=(n+1-j)x^{n+j}-\frac{1}{2}\sum_{k=1}^{j-1}x^{n+k}+\frac{1}{2}\sum_{k=j+1}^{n}x^{n+k}\nonumber\\
&\quad+\frac{t}{2}\left((x^{n+j})^2+e^{(x^{j-1}-x^j)}+e^{(x^j-x^{j+1})}\right) \, ,\\
\eta^{n+j}_{(1)}&=\frac{1}{2}(x^{n+j})^2+(n+2-j)e^{(x^{j-1}-x^j)}-(n-j)e^{(x^j-x^{j+1})}+\nonumber\\
&\quad +\frac{t}{2}(x^{n+j-1}+x^{n+j})e^{(x^{j-1}-x^j)}-\frac{t}{2}(x^{n+j}+x^{n+j+1})e^{(x^j-x^{j+1})}\, ,
}\end{equation}
where conventions \eref{conventions} must be used.

It is straightforward to prove that $\eta_{(2)}=(\eta^j_{(2)},\eta^{n+j}_{(2)})$ with $j=1,.....,n$ and
\begin{equation}\label{eta2}
\eqalign{\eta^j_{(2)}&=j-\frac{t}{2}x^{n+j}\, ,\\
\eta^{n+j}_{(2)}&=-\frac{1}{2}x^{n+j}-\frac{t}{2}\left(e^{(x^{j-1}-x^j)}-e^{(x^j-x^{j+1})}\right)\, ,
}\end{equation}
is a new symmetry vector for the $n$ dimensional Toda Lattice.

The third symmetry vector $\eta_{(3)}=(\eta^j_{(3)},\eta^{n+j}_{(3)})$ with $j=1,.....,n$ is
\begin{equation}\label{eta3}
\eqalign{\eta^j_{(3)}&=t\, ,\\
\eta^{n+j}_{(3)}&=1\, .
}\end{equation}

This symmetry transformation has an interesting feature, it is, in
some sense, the inverse symmetry of $\eta_{(1)}$. In fact, we have
that
\begin{equation}
\sigma^{(0)}=\underset{\eta_{(3)}}{\cal L}\underset{\eta_{(1)}}{\cal L}\sigma^{(0)}\, ,
\end{equation}
where $\sigma^{(0)}$ will be displayed later. The specific details
will be discussed in Appendix 2.

The fourth symmetry vector $\eta_{(4)}=(\eta^j_{(4)},\eta^{n+j}_{(4)})$ with $j=1,.....,n$ is
\begin{equation}\label{eta4}
\eqalign{
\eta^j_{(4)}&=1\, ,\\
\eta^{n+j}_{(4)}&=0\, .
}
\end{equation}

The last and fifth symmetry vector $\eta_{(5)}=(\eta^j_{(5)},\eta^{n+j}_{(5)})$ is
\begin{equation}\label{eta5}
\eqalign{
\eta^j_{(5)}&=\sum_{i=1}^n x^i\, ,\\
\eta^{n+j}_{(5)}&=\sum_{i=1}^n x^{n+i}\, .
}
\end{equation}

%%%%%%%%%%%%%%%%%%%%%%%%%%%%%%%%%%%%%%%%
%%%%%%%%%%%%%%%%%%%%%%%%%%%%%%%%%%%%%%%%

\section{Multi-Lagrangian and multi-Hamiltonian structures}

Now, we will show how the five symmetries allow us to construct new
Lagrangians different from Lagrangian (\ref{lagra1}). Examples of
this Multi-Lagrangian structure in two dimensions are presented.

One way of constructing new Lagrangian structures is described in
Appendix 1.

\subsection{Multi-Lagrangian and multi-Hamiltonian structures associated to symmetry $\eta_{(1)}$}

The $n$ dimensional Lagrangian  given by (\ref{lagra1}) may be
rewritten as ${\hat L}$
\begin{equation}
{\hat L}=\sum_{a=1}^{a=2n}{\hat l}_a{\dot x}^a+{\hat l}_0\, ,
\label{lhat}
\end{equation}
where ${\hat l}_j=x^{n+j}$, ${\hat l}_{j+n}=0$ and
\begin{equation}
{\hat l}_0=-\frac{1}{2}\sum_{j=1}^n(x^{j+n})^2-\sum_{j=1}^n e^{x^j-x^{j+1}}\, ,
\end{equation}
with $j=1,......,n$.

We  now define a new Lagrangian $L$
by adding a total time derivative of a function $\lambda$ to Lagrangian ${\hat L}$
such that the Lagrangian one-form ${l^{(0)}}_a$, defined below, satisfies the Master Equation,
\begin{equation}
L={\hat L}+ \frac{d\lambda}{dt}\, ,
\label{laggg0}
\end{equation}
which may be also be written as ($a=1,...,2n$)
\begin{equation}
L={l^{(0)}}_a(\dot x^a-f^a)\, .
\label{laggg1}
\end{equation}

The Lagrangian one-form ${l^{(0)}}_a$ is defined by
\begin{equation}
{l^{(0)}}_a={\hat l}_a+\lambda,_{a}\, ,
\label{l0nd2}
\end{equation}
where the $\lambda$ function must satisfy
\begin{equation}
\frac{\partial \lambda}{\partial x^a}f^a+\frac{\partial \lambda}{\partial t}=-\frac{1}{2}\sum_{j=1}^n(x^{j+n})^2+\sum_{j=1}^n e^{x^j-x^{j+1}}\, .
\label{funcionlam}
\end{equation}

One solution for $\lambda$ satisfying \Eref{funcionlam} is
\begin{equation}
\lambda=-\frac{t}{2}\sum_{j=1}^n(x^{j+n})^2-t\sum_{j=1}^n e^{x^j-x^{j+1}}+\sum_{j=1}^n (2j-1)x^{j+n}\, ,
\label{funcionayuda}
\end{equation}
and then using eq.\eref{l0nd2} we have
\begin{equation}\label{l0nd}
\eqalign{
{l^{(0)}}_j&=x^{n+j}+t\left(e^{(x^{j-1}-x^j)}-e^{(x^j-x^{j+1})}\right)\, ,\\
{l^{(0)}}_{n+j}&=-tx^{n+j}+(2j-1)\, .
}
\end{equation}

On the other hand, from Lagrangian \eref{laggg1} we get the canonical momenta
\begin{equation}
{p^{(0)}}_j=\frac{\partial L}{\partial \dot x^j}={l^{(0)}}_j\, .
\end{equation}

Besides, we construct the 2n-dimensional matrix ${\sigma^{(0)}}_{ab}={l^{(0)}}_{a, b}-{l^{(0)}}_{b, a}$
\begin{equation}
{\sigma^{(0)}}_{ab}=\left(\begin{array}{cc}
{\mathbf 0}_{N\times N}&\mathbf{1}_{N\times N}\\
-\mathbf{1}_{N\times N}&{\mathbf 0}_{N\times N}
\end{array}\right)\, .
\label{sigma0nd}
\end{equation}

Using $l^{(0)}$ and matrix \eref{sigma0nd} we will get the Hamiltonian given by \Eref{hamil1}. We can construct a Lagrangian one-form $l^{(1)}$ that satisfies the Master Equation as
\begin{equation}
{l^{(1)}}_a=\underset{\eta_{(1)}}{\cal L}\ {l^{(0)}}_a\, ,
\end{equation}
where $\eta_{(1)}$ is the generalized Chavchanidze's symmetry given by eq.\eref{eta1}. Then, the new Lagragian $L^{(1)}$ is
\begin{equation}
L^{(1)}={l^{(1)}}_a\left({\dot x}^a-f^a\right)={l^{(1)}}_a{\dot x}^a+l_0^{(1)}\, ,
\end{equation}
with $l_0^{(1)}=-{l^{(1)}}_a f^a$ ($a=1,...,2n$).

We can construct a new Hamiltonian by using the equations of motion.
The Lagrangian one-form ${l^{(1)}}_a$ gives rise to a new matrix
${\sigma^{(1)}}_{ab}={l^{(1)}}_{a,b}-{l^{(1)}}_{b,a}$. It can be
proved that, when $\frac{\partial}{\partial t}{\sigma^{(1)}}_{ab}=0$
\cite{sah2} the equations of motion are
\begin{equation}
{\sigma^{(1)}}_{ab}{\dot x}^b=l_{0\ ,a}^{(1)}-{l^{(1)}}_{a,0}=-\frac{\partial H^{(1)}}{\partial x^a}\, .
\label{eqsigma}
\end{equation}

We could, in principle, find the Hamiltonian $H^{(1)}$ using our knowledge of ${l^{(1)}}_a$ and $l_0^{(1)}$. We can iterate this procedure to find a new Lagrangian one-form as ${l^{(2)}}_a=\underset{\eta_{(1)}}{\cal L} {l^{(1)}}_a$, a new matrix ${\sigma^{(2)}}_{ab}={l^{(2)}}_{a,b}-{l^{(2)}}_{b,a}$ and a new Hamiltonian $H^{(2)}$ that give rise to the same equations of motion
\begin{equation}
{\sigma^{(2)}}_{ab}{\dot x}^b=l_{0\ ,a}^{(2)}-{l^{(2)}}_{a,0}=-{H^{(2)}},_a\, .
\label{eqsigma2}
\end{equation}

Proceeding in the same fashion, we can construct a Multi-Lagrangian system.

Let's see an example in two dimensions ($n=2$). With two particles,
the first order Toda model Lagrangian $\hat L =l_a{\dot x}^a+l_0$ is
\begin{equation}
\hat L=x^3{\dot x}^1+x^4{\dot x}^2-\left(\frac{1}{2}(x^3)^2+\frac{1}{2}(x^4)^2+e^{x^1-x^2}\right)\, ,
\label{lag2d}
\end{equation}
and the Euler-Lagrange equations are
\begin{equation}
\eqalign{{\dot x}^1&=x^3=f^1\, ,\\
{\dot x}^2&=x^4=f^2\, ,\\
{\dot x}^3&=-e^{x^1-x^2}=f^3\, ,\\
{\dot x}^4&=e^{x^1-x^2}=f^4\, .
}\label{ecmotion2d}
\end{equation}

We can write the Lagrangian \eref{lag2d} as
\begin{equation}
L^{(0)}={l^{(0)}}_a({\dot x}^a-f^a)
\label{lag2d2}
\end{equation}
where
\begin{equation}
\eqalign{
l^{(0)}&=\left({l^{(0)}}_1, {l^{(0)}}_2, {l^{(0)}}_3, {l^{(0)}}_4\right)\, ,\\
&=(x^3-te^{x^1-x^2},\ x^4+te^{x^1-x^2},\ 1-t x^3,\ 3-tx^4 )\, .
}\end{equation}
The Lagragian \eref{lag2d2} is equivalent to \eref{lag2d} because they produce the same equations of motion. The momenta will be
\begin{equation}
\eqalign{
{p^{(0)}}_{1}&=\frac{\partial L^{(0)}}{\partial \dot x^1}={l^{(0)}}_1\, ,\\
{p^{(0)}}_{2}&=\frac{\partial L^{(0)}}{\partial \dot x^2}={l^{(0)}}_2\, .
}
\end{equation}

We can construct the matrix
\begin{equation}
{\sigma^{(0)}}_{ab}={l^{(0)}}_{a,b}-{l^{(0)}}_{b,a}=\left(\begin{array}{cccc}
0&0&1&0\\
0&0&0&1\\
-1&0&0&0\\
0&-1&0&0
\end{array}\right)\, ,
\label{sigma2d}
\end{equation}
and, if we define $l_0^{(0)}=-{l^{(0)}}_a f^a$, then the equations of motion will be
\begin{equation}
{H^{(0)}}_{,a}={l^{(0)}}_{a,0}-{l_0^{(0)}}_{,a}\, ,
\end{equation}
and from here, it is possible to calculate the first Hamiltonian
\begin{equation}
H^{(0)}=\frac{1}{2}(x^3)^2+\frac{1}{2}(x^4)^2+e^{x^1-x^2}\, .
\label{ham2d2}
\end{equation}

Now, we can use the Lie derivative to calculate more Lagrangians. In
this way, we get
\begin{eqnarray}
{l^{(1)}}_a&=&\underset{\eta_{(1)}}{\cal L}{l^{(0)}}_a\nonumber\\
&=& \frac{1}{2}\left((x^3)^2+e^{x^1-x^2}(8-t(x^3+x^4)),\ (x^4)^2+e^{x^1-x^2}(-6+t(x^3+x^4)),\right.\nonumber\\
&&\left. -te^{x^1-x^2}+6x^3-t(x^3)^2-x^4,\ -te^{x^1-x^2}+x^3+8x^4-t(x^4)^2\right)\, ,
\label{balbalbla1}
\end{eqnarray}
with $\eta_{(1)}$ given in \eref{chavsym2d}. Using \Eref{balbalbla1} we can write the second Lagrangian
\begin{equation}
L^{(1)}={l^{(1)}}_a(\dot x^a-f^a)\, .
\label{lagchav2n1}
\end{equation}

The Lagrange brackets matrix is
\begin{equation}
{\sigma^{(1)}}_{ab}={l^{(1)}}_{a,b}-{l^{(1)}}_{b,a}=\left(\begin{array}{cccc}
0&-e^{x^1-x^2}&x^3&0\\
e^{x^1-x^2}&0&0&x^4\\
-x^3&0&0&-1\\
0&-x^4&1&0
\end{array}\right)\,,
\label{sigma1eta1}
\end{equation}
and we can define the Strong Symmetry matrix $\Lambda^{(1)}$ by
\begin{equation}
\Lambda^{(1)}=\sigma^{(1)}\left(\sigma^{(0)}\right)^{-1}=\left(\begin{array}{cccc}
x^3&0&0&e^{x^1-x^2}\\
0&x^4&-e^{x^1-x^2}&0\\
0&-1&x^3&0\\
1&0&0&x^4
\end{array}\right)\, .
\label{strogsy}
\end{equation}

It is worth mentioning that the Master equation for the strong
symmetry matrix $\Lambda$ is equivalent to the Lax equation
\cite{lax}.

Using the Lagrangian \eref{lagchav2n1}, we can construct a new
Hamiltonian for the same system. This second Hamiltonian $H^{(1)}$
is related to $H^{(0)}$ by
\begin{equation}
\frac{\partial H^{(1)}}{\partial x^a}={{\Lambda^{(1)}}_a}^b\ \ \frac{\partial H^{(0)}}{\partial x^b}\, ,
\label{H1H0eta1}
\end{equation}
because $\sigma^{(1)}$ does not depend explicity on time.

Thus, the Hamiltonian that satisfies the \Eref{eqsigma} and \Eref{H1H0eta1}, with $H^{(0)}$ given in \Eref{ham2d2} and $l_0^{(1)}=-{l^{(1)}}_af^a$ is
\begin{equation}
H^{(1)}=\frac{1}{3}\left(\left(x^3\right)^3+\left(x^4\right)^3\right)+\left(x^3+x^4\right)e^{x^1-x^2}\, ,
\label{H1eta1}
\end{equation}
and, if we define the Poisson Brackets ${J^{(1)}}^{ab}=-\left({\sigma^{(1)}}^{ab}\right)^{-1}$ the equations of motion are
\begin{equation}
{\dot x}^a={J^{(1)}}^{ab}\frac{\partial H^{(1)}}{\partial x^b}\, ,
\end{equation}
and they coincide with those of the eq.(\ref{ecmotion2d}).

We construct ${l^{(2)}}_a={\cal L}_{\eta^{(1)}}{l^{(1)}}_a$
\begin{eqnarray}
l^{(2)}&=&\frac{1}{2}\left(-te^{2x^1-2x^2}+(x^3)^3- e^{x^1-x^2}\left(t(x^3)^2+x^4(-13+tx^4)+x^3(-14+tx^4)\right),\right.\nonumber\\
&&\left. te^{2x^1-2x^2}+(x^4)^3+e^{x^1-x^2}\left(t(x^3)^2+x^3(-11+tx^4)+x^4(-10+tx^4)\right),\right.\nonumber\\
&&\left. -(x^3)^2(-11+tx^3)+x^3x^4+(x^4)^2+e^{x^1-x^2}\left(11-t(2x^3+x^4)\right),\right.\nonumber\\
&&\left. (x^3)^2+x^3x^4+(x^4)^2(13-tx^4)+e^{x^1-x^2}\left(13-t(x^3+2x^4)\right)\right)\, .
\label{l2eta12d}
\end{eqnarray}

Thus, the Lagrangian $L^{(2)}$ may be written as
\begin{equation}
L^{(2)}={l^{(2)}}_a(\dot x^a-f^a)\, ,
\end{equation}
and the Lagrange brackets matrix ${\sigma^{(2)}}_{ab}={l^{(2)}}_{a,b}-{l^{(2)}}_{b,a}$ is
\begin{eqnarray*}
\fl {\sigma^{(2)}}_{ab}=\left(\begin{array}{cccc}
0&-\frac{3}{2}e^{x^1-x^2}\left(x^3+x^4\right)&\frac{3}{2}\left(\left(x^3\right)^2+e^{x^1-x^2}\right)&0\\
\frac{3}{2}e^{x^1-x^2}\left(x^3+x^4\right)&0&0&\frac{3}{2}\left(\left(x^4\right)^2+e^{x^1-x^2}\right)\\
-\frac{3}{2}\left(\left(x^3\right)^2+e^{x^1-x^2}\right)&0&0&-\frac{3}{2}\left(x^3+x^4\right)\\
0&-\frac{3}{2}\left(\left(x^4\right)^2+e^{x^1-x^2}\right)&\frac{3}{2}\left(x^3+x^4\right)&0
\end{array}\right)
\end{eqnarray*}
and due to the fact
\begin{equation}
\frac{\partial {\sigma^{(2)}}_{ab}}{\partial t}=0\, ,
\end{equation}
we can construct a third Hamiltonian that satisfies
\begin{equation}
{H^{(2)}}_{,a}={l^{(2)}}_{a,0}-{l_0^{(2)}}_{,a}={{\Lambda^{(2)}}_a}^b\ {H^{(1)}}_{,b}\, ,
\label{H2H1eta1}
\end{equation}
where $l_0^{(2)}=-{l^{(2)}}_af^a$ and $\Lambda^{(2)}=\sigma^{(2)}\left(\sigma^{(1)}\right)^{-1}=\frac{3}{2}\Lambda^{(1)}$. The Hamiltonian that satisfies \Eref{H2H1eta1} is
\begin{eqnarray}
H^{(2)}&=&\frac{3}{4}e^{2x^1-2x^2}\nonumber\\
&&+\frac{3}{2}e^{x^1-x^2}\left(\left(x^3\right)^2+x^3x^4+\left(x^4\right)^2\right)+\frac{3}{8}\left(\left(x^3\right)^4+\left(x^4\right)^4\right)\, ,
\label{H2eta1}
\end{eqnarray}

The Hamiltonian equations associated to $H^{(2)}$ and the Poisson Brackets ${J^{(2)}}^{ab}$ are
\begin{equation}{\dot x}^a={J^{(2)}}^{ab}\frac{\partial H^{(2)}}{\partial x^b}\, ,\end{equation}
and they are identical to those of eq.(\ref{ecmotion2d}), where ${J^{(2)}}=-\left(\sigma^{(2)}\right)^{-1}$.

We may proceed in the same fashion constructing Lagrangian stuctures, because the Lagrange brackets $\sigma$ matrices are time independent. For example, the fourth Lagrangian for two dimensions is
\begin{equation}
L^{(3)}={l^{(3)}}_a(\dot x^a-f^a)\, ,
\end{equation}
where ${l^{(3)}}_a=\underset{\eta^{(1)}}{\cal L} {l^{(2)}}_a$ is
\begin{eqnarray}
l^{(3)}&=&\frac{3}{4}\left((x^3)^4-2e^{2x^1-2x^2}\left(t(x^3+x^4)-9\right)-\right.\nonumber\\
&&\left.-e^{x^1-x^2}\left[t(x^3)^3+(x^3)^2(-20+tx^4)+x^3x^4(tx^4-19)+(x^4)^2(tx^4-18)\right],\right.\nonumber\\
&&\left.(x^4)^4+2e^{2x^1-2x^2}\left(t(x^3+x^4)-8\right)+\right.\nonumber\\
&&\left.+e^{x^1-x^2}\left[t(x^3)^3+(x^3)^2(tx^4-16)+x^3x^4(tx^4-15)+(x^4)^2(tx^4-14)\right],\right.\nonumber\\
&&\left.-e^{2x^1-2x^2}t-\left((x^3)^3(tx^3-16)+(x^3)^2x^4+x^3(x^4)^2+(x^4)^3\right)-\right.\nonumber\\
&&\left.e^{x^1-x^2}\left[3t(x^3)^2+2x^3(tx^4-16)+x^4(tx^4-15)\right],\right.\nonumber\\
&&\left.-e^{2x^1-2x^2}t+(x^3)^3+(x^3)^2x^4+x^3(x^4)^2+(x^4)^3(18-tx^4)-\right.\nonumber\\
&&\left.e^{x^1-x^2}\left[t(x^3)^2+3x^4(tx^4-12)+x^3(2tx^4-19)\right] \right)\, ,
\end{eqnarray}
and the fouth Hamiltonian is
\begin{eqnarray}
H^{(3)}&=&3\left(x^3+x^4\right)e^{2x^1-2x^2}\nonumber\\
&&+3e^{x^1-x^2}\left(x^3+x^4\right)\left(\left(x^3\right)^2+\left(x^4\right)^2\right)\nonumber\\
&&+\frac{3}{5}\left(\left(x^3\right)^5+\left(x^4\right)^5\right)\, .
\label{H3eta1}
\end{eqnarray}

The algorithm may be reiterated to get new Hamiltonians. We have presented four Lagrangians structures which show the richness of the Toda model and the power of the procedure we have adopted.

\subsection{Multi-Lagrangian  and multi-Hamiltonian structures associated to symmetry $\eta_{(2)}$}

It is possible to repeat the same previous analysis for the symmetry $\eta_{2}$ given in \Eref{eta2}. We use $l^{(0)}$ given in \eref{l0nd}, the Lagrangian of \Eref{laggg1} and Lie derivative to construct other Lagrangians. However, this symmetry does not produce a new Hamiltonian structure for the Toda lattice. Each new Hamiltonian will be identical (except for a constant) to Hamiltonian $H^{(0)}$ given by \Eref{hamil1}. In fact, we can construct a Hamiltonian structures appliying Lie derivatives to Hamiltonians. In this way, it can be proved for $n$ dimensions that using $\eta_{(2)}$ we get a $H^{(m)}$ Hamiltonian as
\begin{eqnarray}
H^{(m)}&=&\underset{\eta_{(2)}}{\cal L}\ H^{(m-1)}={\left(\underset{\eta_{(2)}}{\cal L}\right)}^m\ H^{(0)}\nonumber\\
&&=(-1)^mH^{(0)}\, .
\end{eqnarray}

\subsection{Multi-Lagrangian and multi-Hamiltonian structures associated to symmetry $\eta_{(3)}$}

We mentioned earlier that the $\eta_{(3)}$ symmetry given by
\eref{eta3} is related to symmetry $\eta_{(1)}$ in the following
way
\begin{equation}
{\sigma^{(0)}}_{ab}=\underset{\eta_{(3)}}{\cal L}\underset{\eta_{(1)}}{\cal L}{\sigma^{(0)}}_{ab}\, ,
\end{equation}
where ${\sigma^{(0)}}_{ab}$ is given in \eref{sigma2d}.

Thus, if we call upward hierarchy to Lagrangians and Hamiltonians constructed with symmetry $\eta_{(1)}$, we use $\eta_{(3)}$ to construct a downward hierarchy of Lagrangians and Hamiltonians starting from any Lagrangian of the upward hierarchy.

The construction of these new Lagrangians proceeds much in the same way as
it was done in the last two sections. For example, in two
dimensions, we can construct a new $l'^{(1)}$ as
\begin{eqnarray}
{l'^{(1)}}_a&=&\underset{\eta_{(3)}}{\cal L}\ {l^{(2)}}_a\nonumber\\
&=&\frac{3}{2}\left((x^3)^2+e^{x^1-x^2}(9-t(x^3+x^4)),\ (x^4)^2+e^{x^1-x^2}(-7+t(x^3+x^4)),\right.\nonumber\\
&&\left.-te^{x^1-x^2}-x^3(-7+tx^3)+x^4,\ -te^{x^1-x^2}+x^3+x^4(9-tx^4)\right)
\end{eqnarray}
with $l^{(2)}$ given by \Eref{l2eta12d}. The resulting Lagrangian
is
\begin{equation}
L'^{(1)}={l'^{(1)}}_a(\dot x^a-f^a)\, ,
\end{equation}
and the Lagrange brackets and the new Hamiltonian are
\begin{equation}
{\sigma'^{(1)}}_{ab}={l'^{(1)}}_{a,b}-{l'^{(1)}}_{b,a}=3{\sigma^{(1)}}_{ab}\, ,
\end{equation}
\begin{equation}
H'^{(1)}=3H^{(1)}\, ,
\end{equation}
where $\sigma^{(1)}$ is given in \eref{sigma1eta1} and $H^{(1)}$
is given in \eref{H1eta1}. Applying the Lie derivative again, we
can get a new $l'^{(0)}$
\begin{eqnarray}
{l'^{(0)}}_a&=&\underset{\eta_{(3)}}{\cal L}\ {l'^{(1)}}_a\nonumber\\
&=&3\left(x^3-te^{x^1-x^2},\ x^4+te^{x^1-x^2},\ 3-tx^3,\ 5-tx^4\right)\, ,
\end{eqnarray}
giving rise to the Lagrangian
\begin{equation}
L'^{(0)}={l'^{(0)}}_a(\dot x^a-f^a)\, ,
\end{equation}
and $l'^{(0)}$ produces
\begin{equation}
{\sigma'^{(0)}}_{ab}={l'^{(0)}}_{a,b}-{l'^{(0)}}_{b,a}=3{\sigma^{(0)}}_{ab}\, ,
\end{equation}
\begin{equation}
 H'^{(0)}=3H^{(0)}\, ,
\end{equation}
where $\sigma^{(0)}$ is given by \Eref{sigma0nd} and $H^{(0)}$
is defined in \eref{ham2d2}. The $\eta_{(3)}$ symmetry allows us
to reobtain all the multi-Lagrangian and multi-Hamiltonian
structure.

At this point it is not advisable to apply the Lie derivative to
$l'^{(0)}$ because it gives rise to vanishing Lagrange brackets
$\sigma$. To get around this problem we use the inverse matrix to
the Strong Symmetry matrix \eref{strogsy} and $\sigma^{(0)}$ to
construct a Lagrange brackets matrix $\sigma'^{(-1)}$ as
\begin{equation}
{\sigma'^{(-1)}}_{ab}={\left(\Lambda^{(1)}\right)^{-1}}_a^{\ \ c}{\sigma^{(0)}}_{cb}\, ,
\end{equation}
which gives
\begin{eqnarray}
\fl {\sigma'^{(-1)}}_{ab}= \nonumber\\
\fl \left(\begin{array}{cccc}
0&\left(e^{x^2-x^1}x^3x^4-1\right)^{-1}&x^4\left(x^3x^4-e^{x^1-x^2}\right)^{-1}&0\\
-\left(e^{x^2-x^1}x^3x^4-1\right)^{-1}&0&0&x^3\left(x^3x^4-e^{x^1-x^2}\right)^{-1}\\
-x^4\left(x^3x^4-e^{x^1-x^2}\right)^{-1}&0&0&\left(x^3x^4-e^{x^1-x^2}\right)^{-1}\\
0&-x^3\left(x^3x^4-e^{x^1-x^2}\right)^{-1}&-\left(x^3x^4-e^{x^1-x^2}\right)^{-1}&0\nonumber\\
&&
\end{array}\right)\, .\nonumber\\
&&
\label{sigmamenos1}
\end{eqnarray}

Using this matrix, the equations of motion can be written as
\begin{eqnarray}
{\sigma'^{(-1)}}_{ab}\dot x^b&=&\left(\begin{array}{c}0\\0\\-1\\-1\end{array}\right)\nonumber\\
&=&{l^{(-1)}_0}_{,a}-{l^{(-1)}}_{a,0}=-{H^{(-1)}}_{,a}\, ,
\label{ecmotm1}
\end{eqnarray}
where the Hamiltonian is
\begin{equation}
H^{(-1)}=x^3+x^4\, ,
\label{hinvconmo}
\end{equation}

The Hamiltonian \eref{hinvconmo} can be constructed from the
momentum conservation equation. The $l_a^{(-1)}$ that satisfies
\eref{ecmotm1} is
\begin{eqnarray}
l^{(-1)}&=&\left(\frac{x^3x^4}{x^3x^4-e^{x^1-x^2}},\frac{x^3x^4}{x^3x^4-e^{x^1-x^2}},\right.\nonumber\\
&&\left.\frac{x^4}{x^3x^4-e^{x^1-x^2}},\frac{x^3}{-x^3x^4+e^{x^1-x^2}}\right)\, ,
\end{eqnarray}
and $l^{(-1)}_0=-{l^{(-1)}}_af^a$.

We finally get the Lagrangian $L^{(-1)}={l^{(-1)}}_a(\dot x^a-f^a)$.
One keeps getting new Lagrangians by applying Lie derivatives along
this symmetry in the same way as it was done before. This will
produce a new downward Hamiltonian hierarchy.

\subsection{Multi-Lagrangian and multi-Hamiltonian structures associated to
symmetry $\eta_{(4)}$}

The $\eta_{(4)}$ symmetry given in eq.\eref{eta4} behaves in a
different fashion. Consider $l^{(0)}$ defined by \Eref{l0nd}.
Its Lie derivative along $\eta_{(4)}$ vanishes, i.e.,
\begin{equation}
{l^{(1)}}_a=\underset{\eta_{(4)}}{\cal L}{l^{(0)}}_a\equiv 0\, .
\end{equation}

This feature implies that $\eta_{(4)}$ produce Lagrangians and
Hamiltonians which are identically zero. This fact can be seen in a
different way. The Lie derivative of $H^{(0)}$ along $\eta_{(4)}$
vanishes, i.e.,
\begin{equation}
H^{(1)}=\underset{\eta_{(4)}}{\cal L} H^{(0)}\equiv 0\, .
\end{equation}

\subsection{Multi-Lagrangian and multi-Hamiltonian structures associated to symmetry $\eta_{(5)}$}

The symmetry vector $\eta_{(5)}$ given in \eref{eta5} allows us to
find another Lagragian structure for the Toda lattice. For $n$
dimensional lattices, we apply the Lie derivative to $l^{(0)}$ given
in \eref{l0nd}, as we did before, to obtain $l^{(1)}$ as
\begin{equation}\label{l1eta5}
\eqalign{
{l^{(1)}}_j&=2\eta^{n+j}_{(5)}=2\sum_{i=1}^nx^{n+i}\, ,\\
{l^{(1)}}_{n+j}&=-2t\ \eta^{n+j}_{(5)}+n^2\, ,
}
\end{equation}
which gives rise to the first Lagrangian and Hamiltonian structure
obtained by using this symmetry
\begin{equation}
L^{(1)}={l^{(1)}}_a\left(\dot x^a-f^a\right)={l^{(1)}}_a\ \dot x^a-2\left(\sum_{i=1}^nx^{n+i}\right)^2\, ,
\end{equation}
\begin{equation}
H^{(1)}=\left(\sum_{i=1}^nx^{n+i}\right)^2\, .
\end{equation}

This Hamiltonian is the square of the momentum which is a constant
of motion for this problem. If we apply again the Lie derivative, we
get
\begin{equation}\label{l2eta5}
\eqalign{
{l^{(2)}}_j&=4n\ \eta^{n+j}_{(5)}\, ,\\
{l^{(2)}}_{n+j}&=-4nt\ \eta^{n+j}_{(5)}+n^3\, ,
}
\end{equation}
where the second Lagrangian and Hamiltonian are given by
\begin{equation}
L^{(2)}={l^{(2)}}_a\left(\dot x^a-f^a\right)={l^{(2)}}_a\ \dot x^a-4n\left(\sum_{i=1}^nx^{n+i}\right)^2\, ,
\end{equation}
\begin{equation}
H^{(2)}=2n\left(\sum_{i=1}^nx^{n+i}\right)^2\, .
\end{equation}

Applying the same procedure again will produce results which are
similar to what we have already obtained.

%%%%%%%%%%%%%%%%%%%%%%%%%%%%%%%%%%%%%%%%%%%%%%%%%
%%%%%%%%%%%%%%%%%%%%%%%%%%%%%%%%%%%%%%%%%%%%%%%%%%
\section{Algebra of the new symmetries}

The commutator of two vector fields $A^a$ and $B^b$ may be
constructed by defining the operators
\begin{equation}\hat A=A^a\frac{\partial}{\partial x^a},\ \, \ \ \ \ \ \ \ \  \hat B=B^b\frac{\partial}{\partial x^b}\, ,\end{equation}
and their commutator
\begin{eqnarray}
[\hat A,\hat B]\phi&=&A^a\frac{\partial}{\partial x^a}\left(B^b\frac{\partial\phi}{\partial x^b}\right)-B^b\frac{\partial}{\partial x^b}\left(A^a\frac{\partial \phi}{\partial x^a}\right)\nonumber\\
&=&\left(A^b\frac{\partial B^a}{\partial x^b}-B^b\frac{\partial A^a}{\partial x^b}\right)\frac{\partial\phi}{\partial x^a}=\left({\cal L}_AB^a\frac{\partial}{\partial x^a}\right)\phi\, .
\end{eqnarray}

In this way, the commutator of two vector fields may be defined by
\begin{equation}
 [A,B]={\cal L}_AB= -{\cal L}_BA\, .
\end{equation}

We can now compute the $a$ component of the commutators between any
two symmetry vector fields $\eta_{(m)}$ and $\eta_{(n)}$ with
$m,n=1,2,3,4$
\begin{eqnarray}
 [\eta_{(m)},\eta_{(n)}]^a&=&{\cal L}_{\eta_{(m)}}\eta^a_{(n)}\nonumber\\
&=&\frac{\partial\eta^a_{(n)}}{\partial x^i}\eta^i_{(m)}+\frac{\partial\eta^a_{(n)}}{\partial x^{n+i}}\eta^{n+i}_{(m)}\nonumber\\
&&-\eta^i_{(n)}\frac{\partial\eta^a_{(m)}}{\partial x^i}-\eta^{n+i}_{(n)}\frac{\partial\eta^a_{(m)}}{\partial x^{n+i}}
\end{eqnarray}
with $a=1,...,2n$ and $i=1,...,n$. After some algebra it is easy to
prove that the new symmetries define a Lie algebra given by the
following commutation relations (all other commutators vanish)
\begin{equation}\label{etaconmutador}
\eqalign{
[\eta_{(1)},\eta_{(2)}]^a&=\frac{1}{2}\eta^a_{(1)}\, ,\\
[\eta_{(2)},\eta_{(3)}]^a&=\frac{1}{2}\eta^a_{(3)}\, ,\\
[\eta_{(3)},\eta_{(1)}]^a&=\frac{3}{2}(n+1)\eta^a_{(4)}-2\eta^a_{(2)}\, ,\\
[\eta_{(3)},\eta_{(5)}]^a&=n\ \eta^a_{(3)}\, ,\\
[\eta_{(4)},\eta_{(5)}]^a&=n\ \eta^a_{(4)}\, ,\\
[\eta_{(2)},\eta_{(5)}]^a&=\frac{1}{2}n(n+1)\eta^a_{(4)}\, ,\\
[\eta_{(1)},\eta_{(5)}]^a&=n\ \eta^j_{(1)}\eta^a_{(4)}+n\ \eta^{j+n}_{(1)}\left(1-\eta^a_{(4)}\right)\nonumber\\
&\quad+\eta^{n+j}_{(5)}\left(2\eta^a_{(2)}-\frac{3}{2}(n+1)\eta^a_{(4)}\right)\, ,
}
\end{equation}
with $j$ any number such $1\leq j\leq n$.

%%%%%%%%%%%%%%%%%%%%%%%%%%%%%%%%%%%%%%%%

\section{Conclusions}

We have explicitly presented five different new symmetries for the
dynamics defined by the n dimensional Toda lattice and the first
order Lagrangians, Hamiltonians as well as the corresponding
Lagrange and Strong Symmetries associated to
each of them. Moreover, we showed that the commutators of the
symmetry vector fields give rise to a Lie algebra.

A multi-Lagrangian structure is obtained by taking the Lie
derivative of one-form Lagrangians along each of the symmetry vector
fields. These Lagrangians give rise to equivalent (although not
identical) equations of motion. (They are not identical because
their Lagrange bracket matrices are different). By the same token,
we presented Hamiltonian structures which are different from each other but are
nevertheless equivalent.

%%%%%%%%%%%%%%%%%%%%%%%%%%%%%%%%%%%%%%%%%%%%%%%%%%%%%%
\ack

F A A is very grateful to Programa MECE Educaci\'on Superior for the
Doctoral Scholarship UCH0008.

\appendix{
\section{}

In the inverse problem of the calculus of variations in first order \cite{proinv}, we have a Lagrangian as $L=L(q^i, {\dot q}^j,t)$ where we can define the velovity variable $u^j$ as
\begin{equation*}
u^j={\dot q}^j\, ,
\end{equation*}
and we can define a new Lagrangian $\bar L=L(q^i, u^j,t)$ using these variables.

We define a first order Lagrangian as \cite{sah2}
\begin{equation*}
\bar{\bar L}=\frac{\partial{\bar L}}{\partial u^j}\left({\dot q}^j-u^j\right)+{\bar L}(q^i, u^j, t)\, .
\end{equation*}

Using this Lagrangian, the equations of motion for $u^j$ are
\begin{equation*}
\frac{\partial^2{\bar L}}{\partial u^i\partial u^j}\left({\dot q}^j-u^j\right)-\frac{\partial \bar{L}}{\partial u^i}+\frac{\partial {\bar L}}{\partial u^i}=0\, .
\end{equation*}

If the matrix $\frac{\partial^2{\bar L}}{\partial u^i\partial u^j}$
is regular, these equations are equivalent to our definition
\begin{equation*}
{\dot q}^j-u^j=0\, .
\end{equation*}

The equations of motion for $q^j$
\begin{equation*}
\frac{\partial^2\bar{L}}{\partial q^i\partial u^j}\left({\dot q}^j-u^j\right)-\frac{d}{dt}\left(\frac{\partial {\bar L}}{\partial u^i}\right)+\frac{\partial {\bar L}}{\partial q^i}=0\, ,
\end{equation*}
the first term is zero because ${\dot q}^j=u^j$, and the equations
of motion finally are
\begin{equation*}
-\frac{d}{dt}\left(\frac{\partial {\bar L}}{\partial {\dot q}^i}\right)+\frac{\partial {\bar L}}{\partial q^i}=0\, .
\end{equation*}

We can say that Lagrangian at first order $\bar{\bar L}$ is equivalent to Lagrangian at second order $L$. We can write $\bar{\bar L}$ as ($a=1,.....,2n$)
\begin{equation}
\bar {\bar L}=l_a{\dot x}^a+l_0\, ,
\label{la}
\end{equation}
where
\begin{eqnarray*}
\eqalign{
x^i=q^i\, ,\nonumber \\
x^{j+n}=u^j\, ,\nonumber\\
l_i=\frac{\partial \bar L}{\partial u^i}\, ,\nonumber \\
l_{j+n}=0\, ,\nonumber\\
l_0=-\frac{\partial \bar L}{\partial u^i}u^i+\bar L\, ,
}\end{eqnarray*}
with $i,j=1,...,n$. Now, we can get the equations of motion for
$x^a$ as
\begin{equation*}
\sigma_{ab}{\dot x}^b+\frac{\partial l_a}{\partial t}-\frac{\partial l_0}{\partial x^a}=0\, ,
\end{equation*}
where $\sigma_{ab}=\frac{\partial l_a}{\partial x^b}-\frac{\partial
l_b}{\partial x^a}$. If $\det \sigma\neq 0$, then there is a matrix
$J^{ab}$ such $J^{ab}\sigma_{bc}=-\delta^a_c$ and
\begin{equation*}
{\dot x}^a=J^{ab}\left(\frac{\partial l_b}{\partial t}-\frac{\partial l_0}{\partial x^b}\right)\, .
\end{equation*}

If we demand that curl of $\frac{\partial l_b}{\partial t}-\frac{\partial l_0}{\partial x^b}$ be zero, then $\sigma$ does not depend on time, and
\begin{equation*}
\frac{\partial}{\partial x^a}\left(\frac{\partial l_b}{\partial t}-\frac{\partial l_0}{\partial x^b}\right)-\frac{\partial}{\partial x^b}\left(\frac{\partial l_a}{\partial t}-\frac{\partial l_0}{\partial x^b}\right)=0\, ,
\end{equation*}
and this implies
\begin{equation*}
\frac{\partial^2 l_b}{\partial x^a\partial t}-\frac{\partial^2 l_a}{\partial x^b\partial t}=0\, ,
\end{equation*}
or
\begin{equation*}
\frac{\partial\sigma_{ab}}{\partial t}=0\, ,
\end{equation*}
and this means
\begin{equation*}
\frac{\partial J^{ab}}{\partial t}=0\, .
\end{equation*}

If $\sigma$ does not depend on time, we can always find a function
such that its gradient is $\frac{\partial l_b}{\partial
t}-\frac{\partial l_0}{\partial x^b}$. This is the Hamiltonian
function such that
\begin{equation}
\frac{\partial H}{\partial x^b}=\frac{\partial l_b}{\partial t}-\frac{\partial l_0}{\partial x^b}
\label{way1}
\end{equation}
and the motion equations are now the (first order) Hamilton
equations
\begin{equation*}
{\dot x}^a=J^{ab}\frac{\partial H}{\partial x^b}=f^a\, .
\end{equation*}

Sometimes, $l_a$ of \Eref{la} does not satisfy the Master
equation. We can now construct a new $\bar l_a$ (which satisfies the
Master equation) by
\begin{equation*}
\bar l_a={\cal L}_\eta l_a\, ,
\end{equation*}
such that now the Lagrangian can be written as
\begin{equation*}
L=\bar l_a\left({\dot x}^a-f^a\right)\, .
\end{equation*}

The new $\sigma$ matrix is
\begin{equation*}
\bar\sigma_{ab}=\frac{\partial \bar l_a}{\partial x^b}-\frac{\partial \bar l_b}{\partial x^a}\, .
\end{equation*}

This construction of $\bar\sigma$ is equivalent to the definition
$\bar\sigma_{ab}={\cal L}_\eta\sigma_{ab}$. Using Lie derivatives it
is possible to construct many Hamiltonian functions computing many
$l_a$ and using them in conjunction with \Eref{way1}.

Furthermore, we can define a new matrix $\Lambda$ as
\begin{equation*}
{\Lambda_a}^b=\bar\sigma_{ac}\left(\sigma^{-1}\right)^{cb}\, ,
\end{equation*}
or, in other way, $\bar \sigma_{ab}={\Lambda_a}^c\sigma_{cb}$ and $\bar J^{ab}=J^{ac}{\left(\Lambda^{-1}\right)_c}^b$.

If we have more than one Hamiltonian, say $H$ and $\bar H$, both must satisfy the equation
\begin{eqnarray*}
{\dot x}^a=J^{ab}\frac{\partial H}{\partial x^b}={\bar J}^{ab}\frac{\partial \bar H}{\partial x^b}={J}^{ac}{\left(\Lambda^{-1}\right)_c}^b\frac{\partial \bar H}{\partial x^b}\, ,
\end{eqnarray*}
then, we can find a relation between $H$ and $\bar H$
\begin{equation}
\frac{\partial \bar H}{\partial x^a}={\Lambda_a}^b\frac{\partial H}{\partial x^b}\, .
\label{way2}
\end{equation}

It is possible sometimes to construct other Hamiltonian functions
using \eref{way2} with the help of $\Lambda$.

%%%%%%%%%%%%%%%%%%%%%%%%%%%%%%%%%%%%%%%%%%%%%%%%
\section{}

Symmetries $\eta_{(1)}$, $\eta_{(2)}$ and $\eta_{(5)}$ may be found
by solving the Master equation. The symmetry $\eta_{(4)}$ is the
simplest non-trivial case of a symmetry that satisfies the Master
equation.
However, symmetry $\eta_{(3)}$ was found in a different fashion. We
were looking for a symmetry that was the "inverse" one of
$\eta_{(1)}$.

We have $\sigma^{(0)}$ given by \eref{sigma0nd} and we can define
the $\sigma'$ matrix like $\sigma'={\cal
L}_{\eta_{(1)}}\sigma^{(0)}$. Let's assume that a symmetry $\eta'$
such that $\sigma^{(0)}={\cal L}_{\eta'}\sigma'$ exists.
In other words, the $\eta'$ symmetry must fulfill
\begin{equation*}
\sigma^{(0)}={\cal L}_{\eta'}{\cal L}_{\eta_{(1)}}\sigma^{(0)}\, .
\end{equation*}

We can write this equation in an extended form as
\begin{eqnarray}
\sigma^{(0)}_{ab}&=&\left[\frac{\partial^2\sigma^{(0)}_{ab}}{\partial x^c\partial x^d}\eta^d_{(1)}\eta'^c+\left(\frac{\partial\sigma^{(0)}_{ab}}{\partial x^d}\frac{\partial\eta^d_{(1)}}{\partial x^c}+\frac{\partial\sigma^{(0)}_{db}}{\partial x^c}\frac{\partial\eta^d_{(1)}}{\partial x^a}+ \frac{\partial\sigma^{(0)}_{ad}}{\partial x^c}\frac{\partial\eta^d_{(1)}}{\partial x^b}\right)\eta'^c\right.\nonumber\\
&&\left.+\frac{\partial\sigma^{(0)}_{cb}}{\partial x^d}\eta^d_{(1)}\frac{\partial\eta'^c}{\partial x^a}+\frac{\partial\sigma^{(0)}_{ac}}{\partial x^d}\eta^d_{(1)}\frac{\partial\eta'^c}{\partial x^b}\right]\nonumber\\
&&+\sigma^{(0)}_{db}\frac{\partial^2\eta^d_{(1)}}{\partial x^c\partial x^a}\eta'^c+\sigma^{(0)}_{ad}\frac{\partial^2\eta^d_{(1)}}{\partial x^c\partial x^b}\eta'^c+\left(\sigma^{(0)}_{cd}\frac{\partial\eta^d_{(1)}}{\partial x^b}+\sigma^{(0)}_{db}\frac{\partial\eta^d_{(1)}}{\partial x^c}\right)\frac{\partial\eta'^c}{\partial x^a}\nonumber\\
&&+\left(\sigma^{(0)}_{ad}\frac{\partial\eta^d_{(1)}}{\partial
x^c}+\sigma^{(0)}_{dc}\frac{\partial\eta^d_{(1)}}{\partial
x^a}\right)\frac{\partial\eta'^c}{\partial x^b}\, .
\end{eqnarray}

The term in square brackets vanishes in our case, because the
$\sigma^{(0)}$ matrix is coordinate independent. Thus, this equation
reduces to
\begin{eqnarray}
\sigma^{(0)}_{ab}&=&\sigma^{(0)}_{db}\frac{\partial}{\partial x^a}\left(\frac{\partial \eta^d_{(1)}}{\partial x^c}\eta'^c\right)+\sigma^{(0)}_{ad}\frac{\partial}{\partial x^b}\left(\frac{\partial \eta^d_{(1)}}{\partial x^c}\eta'^c\right)\nonumber\\
&&+\sigma^{(0)}_{cd}\left(\frac{\partial\eta^d_{(1)}}{\partial x^b}\frac{\partial\eta'^c}{\partial x^a}-\frac{\partial\eta^d_{(1)}}{\partial x^a}\frac{\partial\eta'^c}{\partial x^b}\right)
\label{siminv1}
\end{eqnarray}

Then, $\eta'$ must satisfy both \Eref{siminv1} and the Master
equation. It is possible to choose a symmetry such that
$\partial\eta'/\partial x=0$, i.e., a symmetry which is time
dependent only. The transformation $\eta_{(3)}$ is such a symmetry
for dimension $n$.

}

\end{document}